# SILICENE STRUCTURES ON SILVER SURFACES

Hanna Enriquez[1], Sébastien Vizzini[2], Abdelkader Kara[3], Boubekeur Lalmi[1,4], and Hamid Oughaddou[1,5,*]

[1]Institut des Sciences Moléculaires d'Orsay, ISMO-CNRS, Bât. 210, Université Paris-Sud, 91405 Orsay-France

[2]IM2NP Faculté des Sciences et Techniques de Marseille, 13397 Marseille, France

[3]Department of Physics, University of Central Florida, Orlando, FL 32816, USA

[4]Synchrotron SOLEIL, F-91192 Gif Sur Yvette, France

[5]Département de Physique, Université de Cergy-Pontoise, 95031 Cergy-Pontoise Cedex, France

Keywords: Silicene, Silicon, Graphene, STM, Nanoribbons, DFT calculations

**Abstract**

In this paper we report on several structures of silicene, the analog of graphene for silicon, on the silver surface Ag(100), Ag(110) and Ag(111). Deposition of Si produces honeycomb structures on these surfaces. In particular, we present an extensive theoretical study of silicene on Ag(111) for which several recent experimental studies were published. Different silicene structures were obtained only by varying the silicon coverage and/or its atomic arrangement. All the studied structures show that silicene is buckled with a Si-Si nearest neighbor distance varying between 2.28 Å and 2.5Å. Due to the buckling in the silicene sheet, the apparent (lateral) Si-Si distance can be as low as 1.89 Å. We also found that for a given coverage and symmetry, one may observe different STM images corresponding to structures that differ by only a translation.

The discovery of graphene and the confirmation of its remarkable properties have encouraged theoreticians to explore the feasibility of silicene [1-7]. They have shown that silicene poses intrinsic stability. However, unlike graphene, the silicene sheets are stable only if a small buckling (0.44 Å) is present [5]. The electronic properties of silicene nano-ribbons and sheets have been found to be similar to those of graphene [4].

Silicene is considered now to be a promising novel material for nano-electronics as it naturally benefits from the vast Si-based R&D (Research and Development) infrastructure. For instance, the fabrication of the electrical contacts, which is an important problem for materials such as graphene, could be facilitated in the case of silicene by using silicides [8]. Even if the silicene is a quite new material and more experiments are mandatory, the obtained results show very promising features that could give a new future to silicon in the electronics industry.

From the experimental point of view the synthesis of silicene remains a very challenging problem, despite the progress of De Crecsenzi [9], and more recently those of S. Al Yamada [10], demonstrating the synthesis of mono and multi-walled silicon nanotubes. Historically, the first thermally stable *Disilene* molecule containing silicon–silicon double bond was synthesized in 1981 [11-12]. In recent experiments of silicon deposition on silver substrates under ultra high vacuum (UHV) we have shown the first evidence of silicene formation with a honeycomb graphene-like structure forming either parallel assembly of one-dimensional nano-ribbons (NRs) or highly ordered sheet of silicene [13-25]. There the internal silicene structure was studied by scanning tunneling microscopy (STM). These results have recently been summarized in the first thorough review on silicene [14].

On the Ag(110) surface, deposition of silicon produces spontaneously silicene NRs oriented along the $[\bar{1}10]$ direction with an internal honeycomb structure (Figure 1a), just like graphene [13]. The honeycombs structure is supported by density functional theory (DFT) calculations [13-15]. Furthermore, using angle resolved photoemission spectroscopy (AR-PES) it has also been shown that the silicene NRs possess a strong metallic character with quantized states in the *s-p* region of the valence band below the Fermi level with a strictly 1D dispersion along the NRs [19]. In addition they exhibit sharp high-resolution PES spectral features [18]. Further, the reactivity of silicene NRs toward oxidation has been investigated [21]. The oxidation process takes place only at the silicene NRs terminations, which indicates that only the NR extremities contain reactive dangling bonds and develops along the $[\bar{1}10]$ direction.

On the Ag(100) surface, deposition of silicon at 230°C induces the formation of silicene NRs on a p(3×3) superstructure [23]. It has previously been shown with atomically resolved STM as well as surface X-ray diffraction (SXRD) method that the p(3×3) structure corresponds to one atomic Si ML made of Si tetramer entities at on top positions of the Ag(100) surface [23]. Beyond 1 Si ML, the silicene NRs start to develop on the p(3×3) (Figure 1b).

On the Ag(111) surface, we have shown that the deposition of 1 Si ML produces a sheet of Si honeycomb structure with a (2√3×2√3)R30° LEED pattern [22]. Figure 1c shows that silicon deposition induces a continuous film of silicene with almost defect-free honeycomb structure.

Very recently four other groups have independently reported ordered phases on the same surface [26-29]. The silicene sheet can present different orientations relatively to the Ag(111) surface by varying the substrate temperature giving rise to different superstructure, (2√3×2√3)R30°, (√13×√13)R13.9° and 4x4 [26-29].

In order to better understand the Si-induced structures on silver surfaces, we performed atomistic simulations of some of the adsorption structures. We concentrate on the adsorption of a silicon layer on Ag(111), the most natural choice to form a honeycomb arrangement of the silicon atoms due to the three-fold symmetry axis this surface orientation has. On this surface, we have studied six different structures, namely the (4×4) cell, two models in the (2√3×2√3) cell, two in the (√13×√13), and one in the (√7×√7) cell. These correspond to the models proposed in Ref. [29], and one additional structural model in the (2√3×2√3) cell where initially all the silicon atoms are laterally located at an equivalent position. We used the DFT approach by solving the Kohn-Sham equations for the electronic structure. We used the Vienna *ab initio* simulation package (VASP) [30]. Exchange-correlation interactions were included with the generalized gradient approximation (GGA) in the Perdew-Burke-Ernzerhof form [31]. The electron-ion interaction was described by the projector augmented wave method [32]. A plane-wave energy cut-off of 250 eV was used. The bulk lattice constant for Ag was found to be 4.154 Å using a k-point mesh of 10×10×10. The slab super-cell approach with periodic boundary conditions was employed to model the surface. The slab consisted of 4 layers of Ag(111). A hexagonal 2D sheet consisting of silicon atoms was adsorbed on the surface according to the structure of interest. The coordinates of all Si atoms and three top-most Ag layers were allowed to relax to the optimum configuration until the forces on every atom coordinate was less than 0.01 eV/Å. Tersoff-Hamann model [33] was used for the simulation of STM images with an *s*-type tip.

The most often observed structure in experiments is the (4×4) cell of the Ag(111) substrate [26-29]. The atomistic model for it is shown in Figure 2, with 18 Si atoms in the unit cell. From figure 2a one can see that the Si hexagons are present in three configurations, either surrounding an hcp or an fcc three-fold hollow site, an on-top, or a bridge site of the first layer of the substrate. As a result, not all the silicon atoms are laying on the silver surface at the same height; rather, their height above the Ag surface is 2.2 Å for the 12 Si atoms that are closer to the surface and 3.0 Å for the six remaining Si atoms. The Ag surface undergoes a vertical corrugation of 0.4 Å. The Si-Si nearest neighbour distance for this case is 2.35 Å, with a lateral distance (would be apparent distance in an STM image if the atoms would appear with similar brightness) of 2.23 Å. The silicene sheet in this (4×4) structure is quite weakly bound to the silver substrate, as the average binding energy per silicon atom is 0.46 eV if we consider all the silicon atoms. Noting that only 12 out of the 18 Si atoms are close to the Ag atoms, this gives 0.70 eV of binding energy per lower-layer Si atoms, still being relatively small. The work function of the surface is practically unchanged upon the adsorption of silicene. The configuration shown in Figure 2 results in a simulated STM filled-states image shown in Figure 3, which is remarkably similar to the observed one [26-29] and agrees with the other simulations [26-27].

In Figure 4 we show the first model in the (2√3×2√3) cell that we denote as (2√3×2√3)-I in the following. There are 18 Si atoms in the cell. The lowest and the highest Si planes are about 2.2 and 3.7 Å above the silver surface, respectively. In this model, the silicene sheet presents a buckling of 1.5 Å, the largest for all the six models presented here. The Si-Si nearest-neighbor distance in this model is the largest one, 2.51 Å, but due to the large buckling the lateral distance is only 1.99 Å. We also found that in this silicene configuration the change, this time an increase, in the work function is 0.47 eV. Due to the large buckling in silicene, the corresponding simulated filled-states STM image (see figure 5) shows only those silicon atoms occupying the highest plane above the silver surface.

The next model we discuss is another configuration corresponding to the (2√3×2√3) structure. Here there are 14 silicon atoms as shown in Figure 6. In this configuration the silicon hexagons surround bridge and on-top sites.

Here again silicon atoms are vertically located in two planes but the majority of the atoms is hosted in the lower plane. These two planes are about 2.2 and 3.2 Å above the silver surface. The top-most Ag layer undergoes a buckling of 0.3 Å. The Si-Si nearest neighbour distance varies between 2.28 and 2.37 Å; with a lateral distance of 2.0 Å, comparable to the one observed by Lalmi *et al* [22]. In contrast with the (2√3×2√3)-I, this configuration did not result in any change in the work function upon silicene adsorption. In Figure 7, we present the corresponding simulated STM image, which is strikingly different from the previous configuration. We see from the analysis of these two (2√3×2√3) configurations that for the same LEED pattern one may see different STM images depending on how the silicene film grew. We argue that several configurations yielding the same LEED pattern may show different STM images because a slight translation of silicene on top of Ag(111) introduce dramatic changes in the STM images without altering the LEED pattern. A very delicate interplay between the surface temperature and flux (probably other factors may enter) may yield a different structure, yet the same periodicity.

Another configuration that was observed by two experimental groups [28-29] is the (√13×√13) surface unit cell. Again we present results for two models, different laterally from each other by a translation. In Figure 8 we show the top and the side views of the first configuration, with 14 Si atoms per unit cell. In this configuration the silicon hexagons surround either a three-fold hollow site, a bridge site or close to an on-top site. From Figure 8, we see again that the silicon atoms occupy two planes above the silver surface with the majority of silicon atoms occupying the lower plane. Note that only one (residing on top of a substrate atom) out of 14 silicon atoms occupies the highest plane. These two planes are 2.2 and 3.4 Å above the silver surface, which itself presents a buckling of about 0.5 Å. The silicon-silicon nearest neighbour distances is either 2.35 Å or 2.43 Å and the lateral buckling is 2.34 Å. In this configuration we obtained a small decrease in the work function by 0.17 eV. The corresponding simulated STM image is shown in Figure 9. Only the atoms (1 out of 14) occupying the highest plane are clearly imaged.

Let us contrast these values with those corresponding to the same structure as the previous one, but differs by a translation of the silicene sheet as shown in figures 8a (top view) and 8b (side view).

Even though the top view of this structure in Figure 10a resembles that of in Figure 8a, the difference is that the silicon atoms occupy two different planes that are much closer to each other than the previous case. In this configuration, four (out of 14) silicon atoms occupy the highest plane, 2.9 Å above the silver surface, and the rest of the silicon atoms occupy a lower plane, 2.1 Å above the surface. The buckling in the top-most Ag layer is 0.4 Å. The nearest neighbor Si-Si distance in this configuration varies between 2.35 and 2.40 Å, with the smallest lateral distance being 1.89 Å. In this configuration, the work function was found to be almost unchanged. In Figure 11 we show the simulated STM image corresponding to the configuration in Figure 10. Again there is a striking difference between the STM images in Figures 9 and 11.

Finally we present results for a further model where the surface unit cell is ($\sqrt{7}\times\sqrt{7}$), as shown in Figure 12a. Note from Figure 12a that the centres of the Si hexagons surround either a bridge or on-top site. Again the silicon atoms occupy two vertical heights above the silver surface at 2.2 and 2.9 Å. Three out of eight silicon atoms occupy the higher plane. The Si-Si nearest-neighbour distances are between 2.32 and 2.34 Å, while the lateral one is 2.2 Å. The buckling in the top-most substrate layer is 0.4 Å. In this configuration a small decrease in the work function was found (0.13 eV). In Figure 13, we present the simulated filled-states STM image corresponding to this structure.

In conclusion, silicene is a two-dimensional material with a honeycomb lattice resembling that of graphene. On Ag(100) and Ag(110), silicene grows as nanoribbons, while on Ag(111) silicene grows as sheets covering very large areas. However, contrary to graphene that can be formed as standalone sheets and ribbons, several groups have only grown silicene on a silver substrate. What one learns from all the published experimental data and the accompanying DFT based calculations is that the formation of silicene on Ag(111) is a much more complicated system than what was claimed in a recent publication by Vogt *et al* where they believed that silicene exists only in the 4x4 structure [26]. Rather, as our calculations and other experimental studies have shown, silicene growth is very complex and its study is still in its debut.

From all the studies presented in this paper we found that the distance deduced from the STM images are a projected one and the buckling measured is that of the charge density, which is in general smaller than the true one. Thus the Si-Si distances in silicene maybe closer to the bulk silicon value (2.35 Å) than previously concluded from STM images alone, which was close to the lateral distance of 2.0 Å. Moreover, we note that in all the cases of silicene on Ag(111) presented here, silicon atoms reside in two layers above the silver substrate, where the lower layer is about 2.2 Å above the silver (buckled) surface.

**Acknowledgements:**

The authors wish to thank Dr A. P. Seitsonen and B. Aufray for fruitful discussions.

**Figure captions :**

**Figure 1:** (a) (21×21 nm$^2$) Filled-states STM image showing straight, parallel 1D Si NRs after deposition of 0.5 Si ML on Ag(110). Upper- left corner, (3.8×4.2 nm$^2$), atomically resolved filled-states STM image revealing honeycomb structure [13]. b) (5×5 nm$^2$) Filled-states STM images on the Ag(001) surface after the deposition of 1.6 Si ML showing the honeycomb structure of the silicene NRs [23]. c) Filled-states STM image (3.8×3.8 nm$^2$) showing honeycomb structure after the deposition of 1 Si ML on Ag(111) [22].

**Figure 2:** Top (a) and side (b) view of the atomistic model of the Si/Ag(111)-(4×4) structure. The silicon atoms are drawn with blue, the top-most Ag layer with orange and the lower Ag layers with grey spheres.

**Figure 3:** Simulated filled-states STM image of the Si/Ag(111)-(4×4) structural model shown in Figure 2 using a bias voltage of -1.4 V.

**Figure 4:** The top (a) and side (b) view of the Si/Ag(111)-(2√3×2√3)-I structural model. The coloring of the atoms is the same as in Figure 3.

**Figure 5:** Simulated filled-states STM image of the Si/Ag(111)-(2√3×2√3)-I structural model shown in Figure 4 using a bias voltage of -1.4 V.

**Figure 6:** Top (a) and side (b) view of the Si/Ag(111)-(2√3×2√3)-II structural model. The coloring of the atoms is the same as in Figure 2.

**Figure 7:** Simulated filled-states STM image of the Si/Ag(111)-(2√3×2√3)-II structural model shown in Figure 6 using a bias voltage of -1.4 V.

**Figure 8:** Top (a) and side (b) view of the Si/Ag(111)-(√13×√13)-I structural model. The coloring of the atoms is the same as in Figure 2.

**Figure 9:** Simulated filled-states STM image of the Si/Ag(111)-(√13×√13)-I structural model shown in Figure 8 using a bias voltage of -1.4 V.

**Figure 10:** Top (a) and side (b) view of the Si/Ag(111)-(√13×√13)-II structural model. The coloring of the atoms is the same as in Figure 2.

**Figure 11:** Simulated filled-states STM image of the Si/Ag(111)-(√13×√13)-II structural model shown in Figure 10 using a bias voltage of -1.4 V.

**Figure 12:** The top (a) and side (b) view of the Si/Ag(111)-(√7×√7) structural model. The coloring of the atoms is the same as in Figure 2.

**Figure 13:** Simulated filled-states STM image of the Si/Ag(111)-(√7×√7) structural model shown in Figure 12 using a bias voltage of -1.4 V.

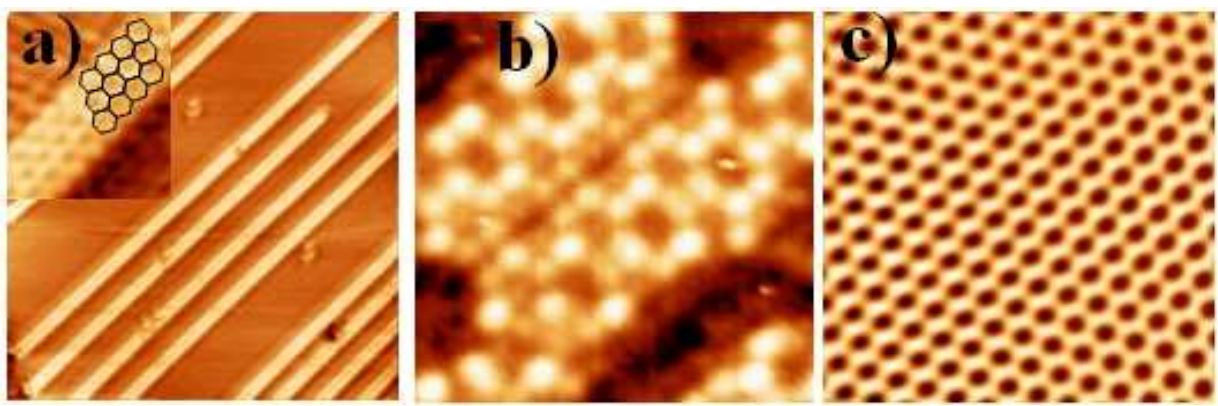

**Figure 1:** (a) (21×21 nm$^2$) Filled-states STM image showing straight, parallel 1D Si NRs after deposition of 0.5 Si ML on Ag(110). Upper-left corner, (3.8×4.2 nm$^2$), atomically resolved filled-states STM image revealing honeycomb structure [13]. b) (5×5 nm$^2$) Filled-states STM images on the Ag(001) surface after the deposition of 1.6 Si ML showing the honeycomb structure of the silicene NRs [23]. c) Filled-states STM image (3.8×3.8 nm$^2$) showing honeycomb structure after the deposition of 1 Si ML on Ag(111) [22].

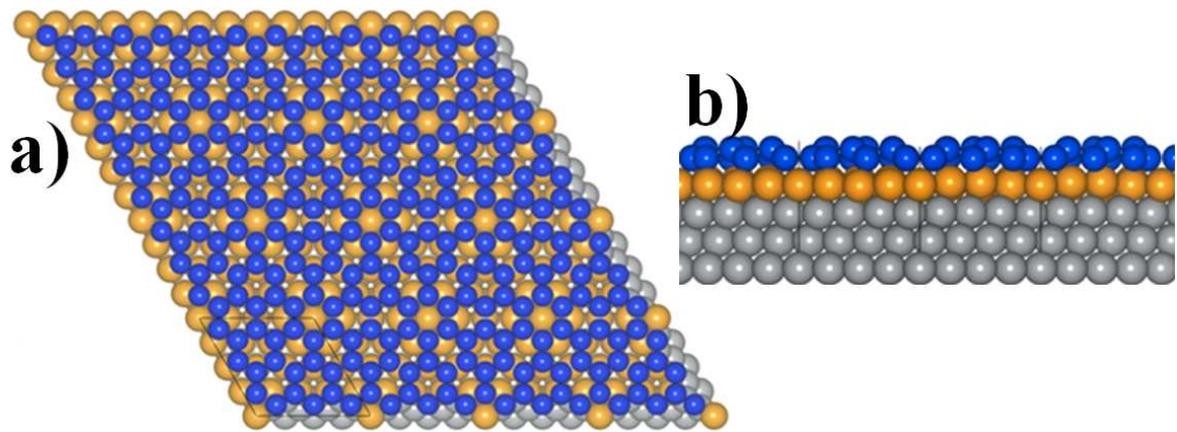

**Figure 2:** Top (a) and side (b) view of the atomistic model of the Si/Ag(111)-(4×4) structure. The silicon atoms are drawn with blue, the top-most Ag layer with orange and the lower Ag layers with grey spheres.

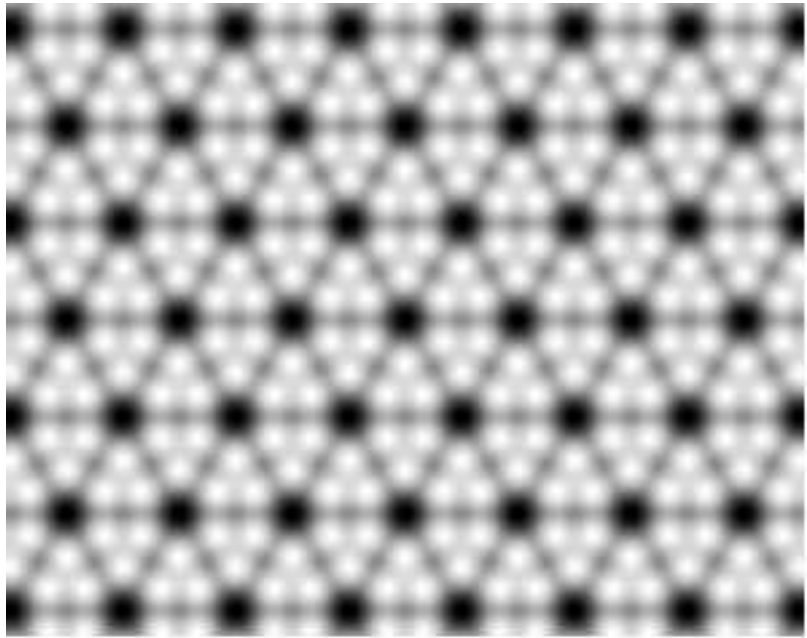

**Figure 3:** Simulated filled-states STM image of the Si/Ag(111)-(4×4) structural model shown in Figure 2 using a bias voltage of -1.4 V.

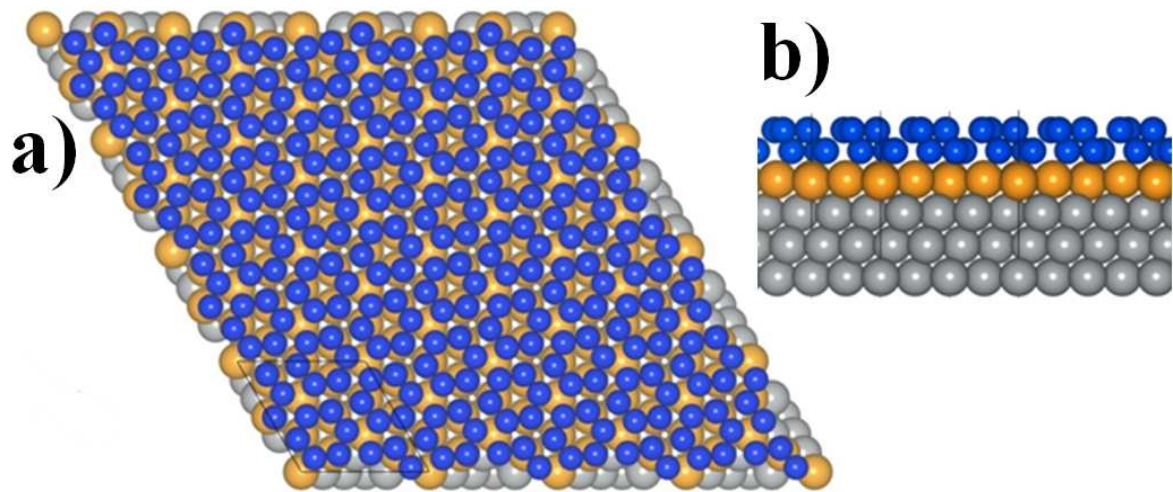

**Figure 4:** The top (a) and side (b) view of the Si/Ag(111)-(2√3×2√3)-I structural model. The coloring of the atoms is the same as in Figure 2.

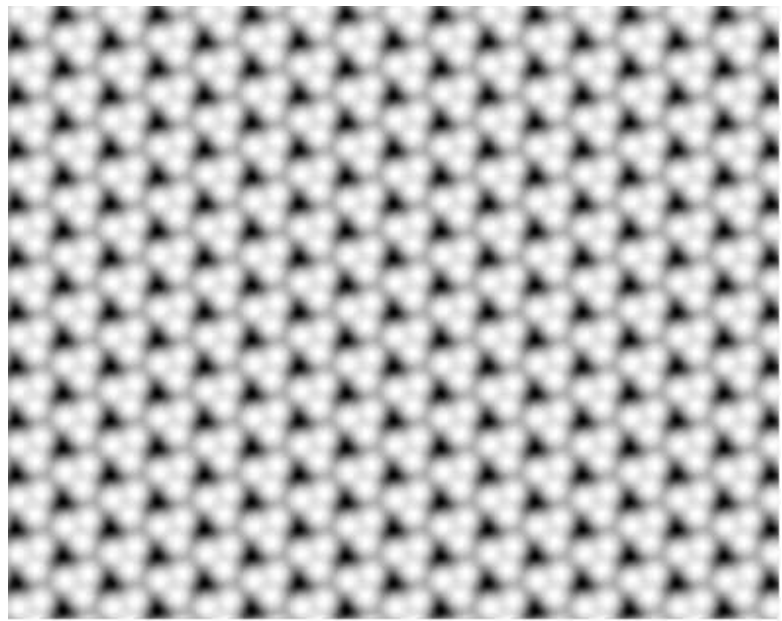

**Figure 5:** Simulated filled-states STM image of the Si/Ag(111)-(2√3×2√3)-I structural model shown in Figure 4 using a bias voltage of -1.4 V.

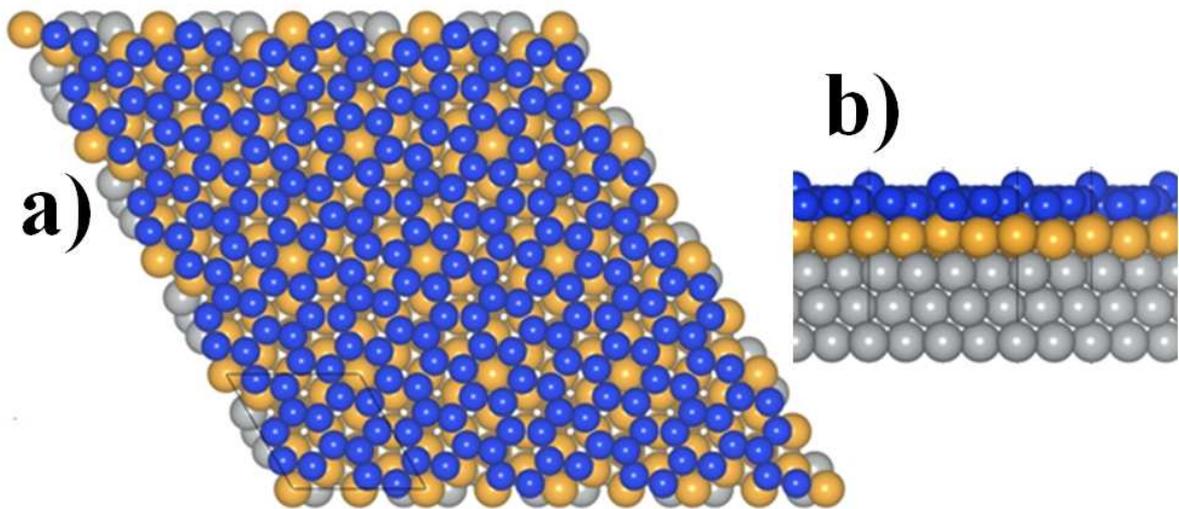

**Figure 6:** Top (a) and side (b) view of the Si/Ag(111)-(2√3×2√3)-II structural model. The coloring of the atoms is the same as in Figure 2.

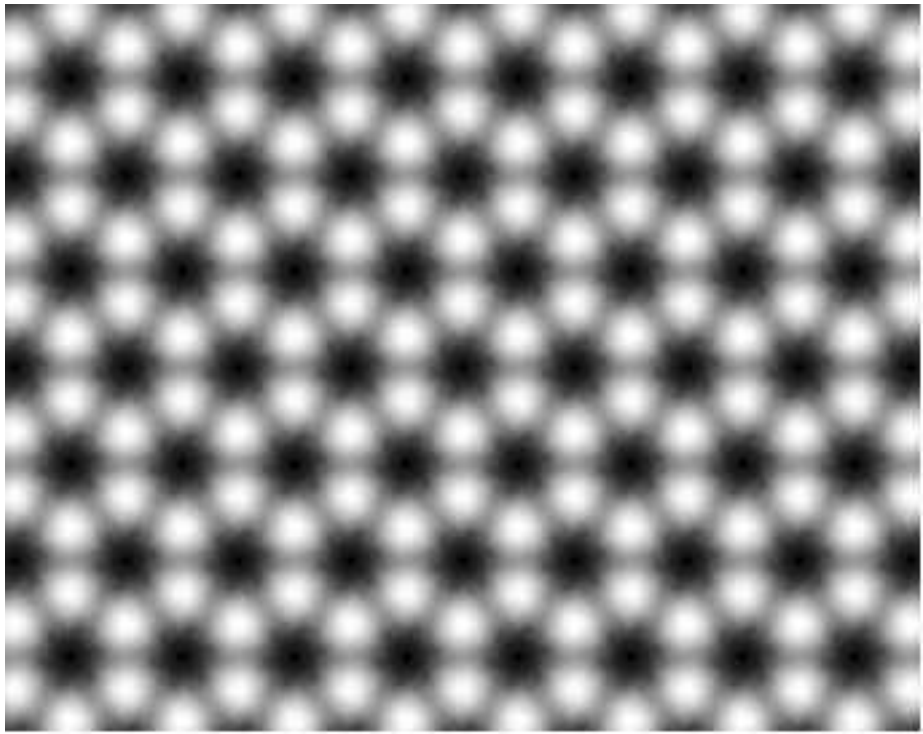

**Figure 7:** Simulated filled-states STM image of the Si/Ag(111)-(2√3×2√3)-II structural model shown in Figure 6 using a bias voltage of -1.4 V.

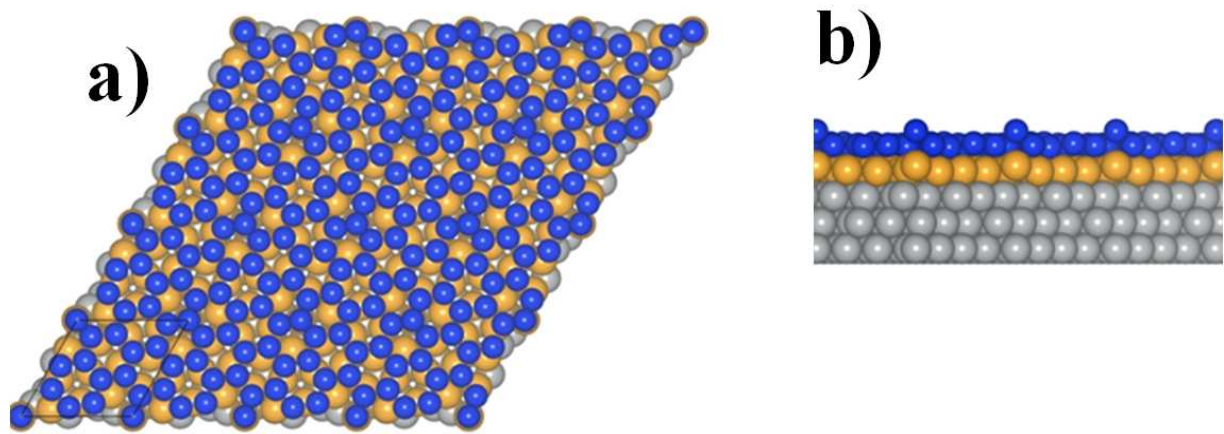

**Figure 8:** Top (a) and side (b) view of the Si/Ag(111)-(√13×√13)-I structural model. The coloring of the atoms is the same as in Figure 2.

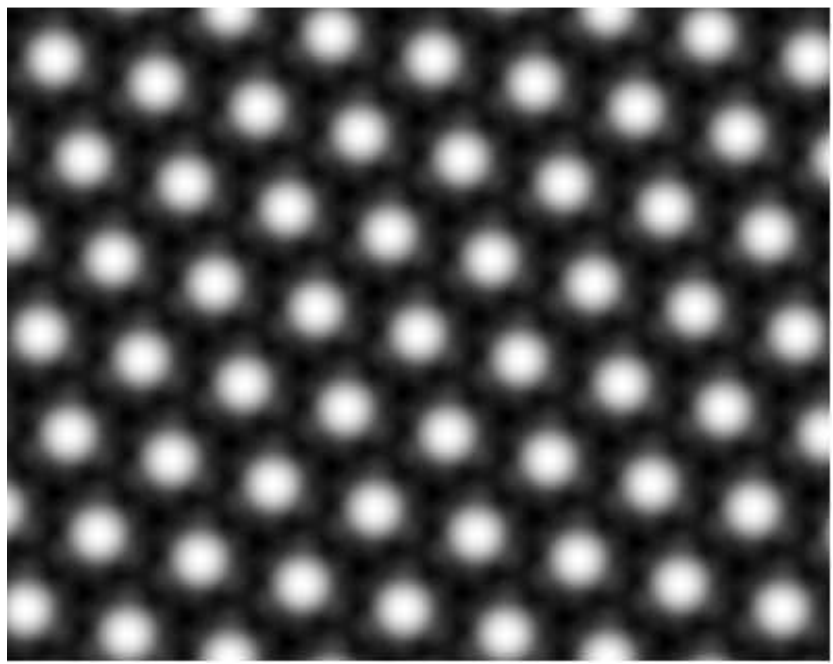

**Figure 9:** Simulated filled-states STM image of the Si/Ag(111)-(√13×√13)-I structural model shown in Figure 8 using a bias voltage of -1.4 V.

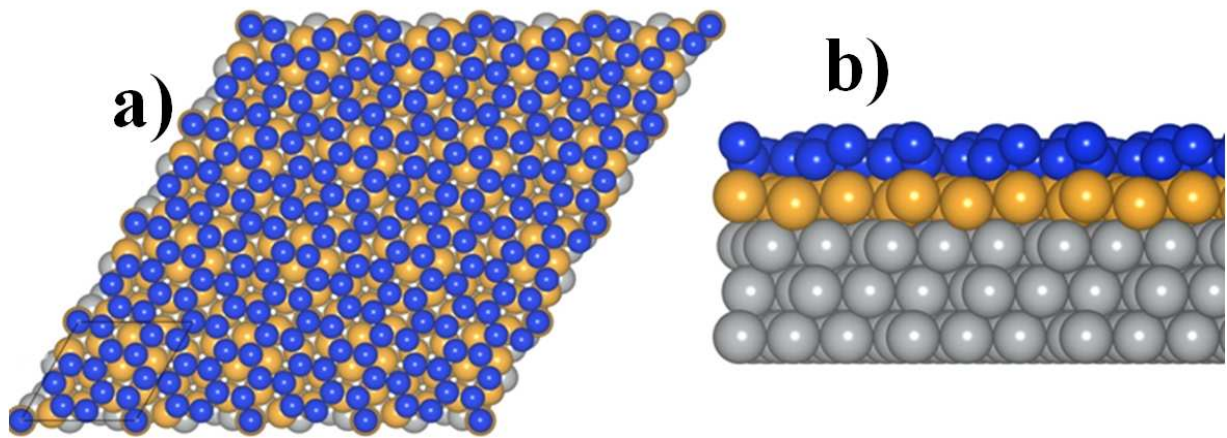

**Figure 10:** Top (a) and side (b) view of the Si/Ag(111)-(√13×√13)-II structural model. The coloring of the atoms is the same as in Figure 2.

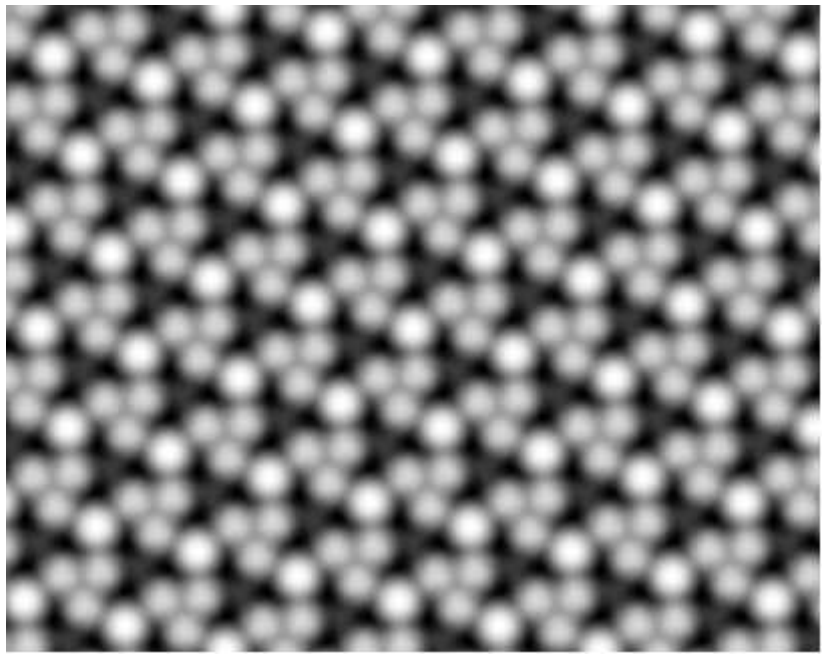

**Figure 11:** Simulated filled-states STM image of the Si/Ag(111)-(√13×√13)-II structural model shown in Figure 10 using a bias voltage of -1.4 V.

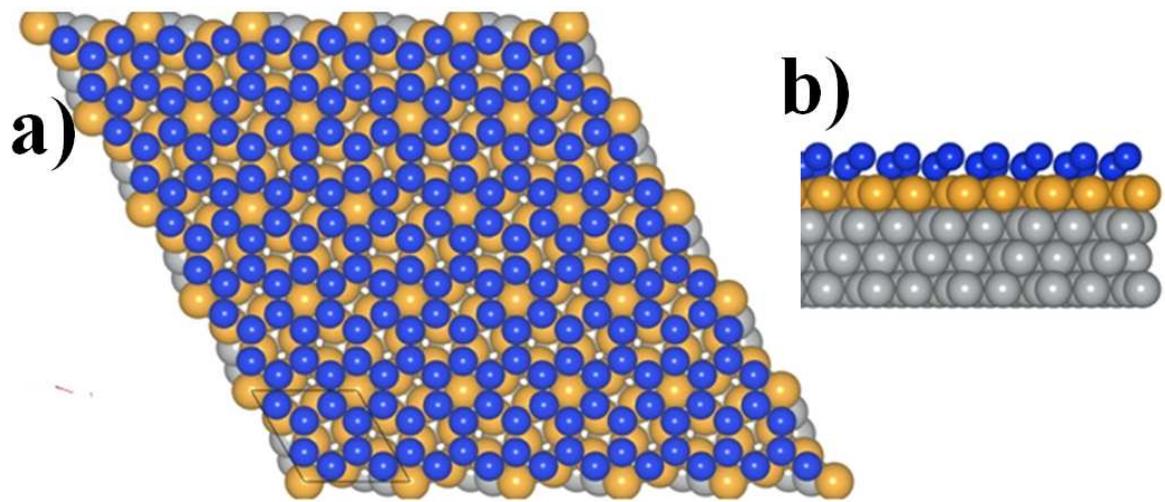

**Figure 12:** The top (a) and side (b) view of the Si/Ag(111)-($\sqrt{7}\times\sqrt{7}$) structural model. The coloring of the atoms is the same as in Figure 2.

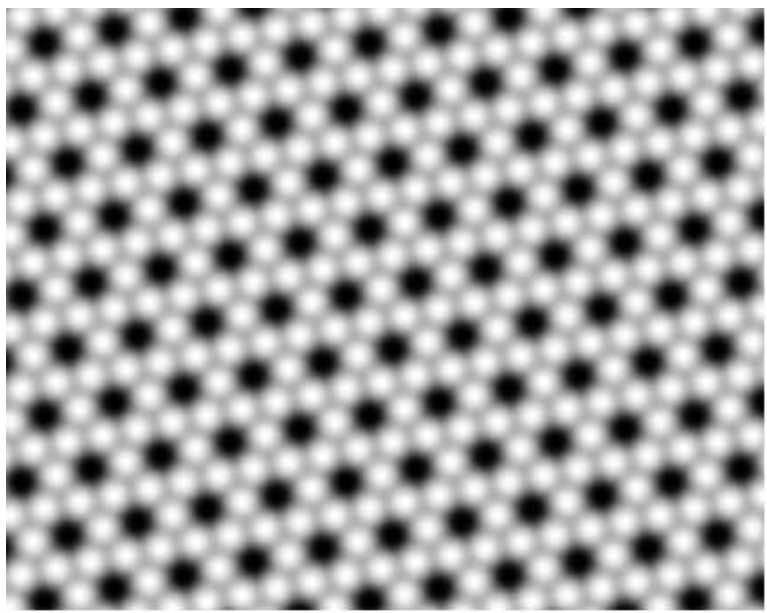

**Figure 13:** Simulated filled-states STM image of the Si/Ag(111)-($\sqrt{7}\times\sqrt{7}$) structural model shown in Figure 12 using a bias voltage of -1.4 V.